\input phyzzx
\input epsf

%
%
%
\catcode`\@=11 
\def\papersize{\hsize=40pc \vsize=53pc \hoffset=0pc \voffset=1pc
   \advance\hoffset by\HOFFSET \advance\voffset by\VOFFSET
   \pagebottomfiller=0pc
   \skip\footins=\bigskipamount \normalspace }
\catcode`\@=12 
\papers

\newcount\figno
\figno=0
\def\fig#1#2#3{
\par\begingroup\parindent=0pt\leftskip=1cm\rightskip=1cm\parindent=0pt
\baselineskip=11pt
\global\advance\figno by 1
\midinsert
\epsfxsize=#3
\centerline{\epsfbox{#2}}
\vskip 12pt
{\bf Fig. \the\figno:} #1\par
\endinsert\endgroup\par
}
\def\figlabel#1{\xdef#1{\the\figno}}
\def\encadremath#1{\vbox{\hrule\hbox{\vrule\kern8pt\vbox{\kern8pt
\hbox{$\displaystyle #1$}\kern8pt}
\kern8pt\vrule}\hrule}}

\vsize=23.cm
\hsize=15.cm

\tolerance=500000
\overfullrule=0pt

\Pubnum={LPTENS-96/09 \cr
{\tt hep-th@xxx/9601129} \cr
January 1996}

\date={}
\pubtype={}
\titlepage
\title{\bf DOES COUPLING TO GRAVITY PRESERVE INTEGRABILITY ?}  
\author{Adel~Bilal}
\address{
CNRS - Laboratoire de Physique Th\'eorique de l'Ecole
Normale Sup\'erieure
\foot{{\rm unit\'e propre du CNRS, associ\'e \`a l'Ecole Normale
Sup\'erieure et l'Universit\'e Paris-Sud}}
 \nextline 24 rue Lhomond, 75231
Paris Cedex 05, France\break
{\tt bilal@physique.ens.fr}
}

\abstract{\noindent
I describe evidence that the (two-dimensional) 
integrable chiral Gross-Neveu model might remain
integrable when coupled to gravity. These are notes based on a lecture given at
the Cargese summer school 1995. The results were obtained in 
collaboration with Ian Kogan.}

\vskip 1.cm
\centerline{\it To appear in the Proceedings of the 1995 Cargese Summer School} 
\centerline{\it ``Low-dimensional Applications of Quantum Field Theory" }

\endpage
\pagenumber=1

\def\PL #1 #2 #3 {Phys.~Lett.~{\bf #1} (#2) #3}
\def\NP #1 #2 #3 {Nucl.~Phys.~{\bf #1} (#2) #3}
\def\PR #1 #2 #3 {Phys.~Rev.~{\bf #1} (#2) #3}
\def\PRL #1 #2 #3 {Phys.~Rev.~Lett.~{\bf #1} (#2) #3}
\def\MPL #1 #2 #3 {Mod.~Phys.~Lett.~{\bf #1} (#2) #3}
\def\d{\partial}
\def\g{\gamma}
\def\m{\mu}
\def\n{\nu}
%
%
%
%
%
%
%
\REF\BK{ A. Bilal and I.I. Kogan, {\it Gravitationally dressed conformal 
field theory and emergence of logarithmic operators}, 
Princeton University preprint PUPT-1482, {\tt hep-th/9407151};\nextline
A. Bilal and I.I. Kogan,
\NP B449 1995 569 , {\tt hep-th/9503209}.}
\REF\GN{ D.J. Gross and A. Neveu, \PR D10 1974 3235 .}
\REF\AL{ N. Andrei and J.H. Lowenstein, \PRL 43 1979 1698 .}
\REF\LOW{ J.H. Lowenstein, {\it Introduction to the Bethe 
ansatz approach in $(1+1)$-dimensional
models} in: Les Houches 1982 (Elsevier).}
\REF\POLY{ A. Polyakov, \MPL A2 1987 893 ;\nextline
V. Knizhnik, A. Polyakov and A. Zamolodchikov, 
\MPL A3 1988 819 .}
\REF\AA{E. Abdalla and M.C.B. Abdalla, {\it Gravitational interactions of integrable
models}, preprint CERN-TH/95-209, {\tt hep-th/9507171}.}

\chapter{Introduction}
     
These notes are based on results obtained in collaboration with Ian Kogan. They have
already been briefly announced in our previous articles [\BK]. Here I will give
some more details.

When a renormalisable quantum field theory is coupled to gravity three things may
happen: the coupled system  is more difficult or even non-renormalisable, it is
simpler, or about the same. The first option is of course the generic one in four
dimensions. In two dimensions, things tend to be better. We know of examples where
adding conformal matter and gravity leads to a topological field theory, i.e. a much
simpler system. In two dimensions, there are also many exactly 
integrable models. What
happens to them when they are coupled to
two-dimensional gravity? To try to get an
idea, Ian Kogan and I have investigated the case of the integrable
chiral Gross-Neveu model. We
did neither prove nor disprove integrability of this model when coupled to gravity, but
we obtained some encouraging indications that integrability might indeed be preserved.

In these notes, after introducing the model, I will describe these indications: we
checked, up to the second non-trivial order whether the two-particle $S$-matrix remains
elastic in the presence of gravity. It turned out that this is the case.

\chapter{The chiral Gross-Neveu model and the Bethe ansatz}

The action of the chiral Gross-Neveu model is given by
$$\eqalign{
S&\sim{1\over\sqrt{2}}\int {\rm d}t {\rm d}x
\left[ -i \bar\psi^j \gamma^a \partial_a \psi^j
+{1\over 2} g^2 (\bar\psi^j\psi^j)^2 -
 {1\over 2} g^2 (\bar\psi^j\gamma_5\psi^j)^2\right]\cr
&=\int {\rm d}t {\rm d}x\left[ i(\psi^j_+)^*\partial_- \psi^j_+
+ i(\psi^j_-)^*\partial_+ \psi^j_-
+\sqrt{2} g^2 (\psi_+^i)^* (\psi_-^j)^*\psi_-^i\psi_+^j \right] \ .\cr}
\eqn\gnaction         
$$
(I will detail my conventions in the next section.)
One has $N$ colours of Dirac fermions
$\psi^j$ ($j=1,\ldots N$), and $\psi_-^j$ and $\psi_+^j$ refer to right and left-moving
components (Weyl fermions).\foot{
In ref. \BK, we mistakenly called $\psi_-$ left-moving and $\psi_+$
right-moving.}
 It is often convenient to rewrite the $\psi$
self-couplings using two auxiliary scalar fields $\sigma$ and $\pi$
with unit propagator and coupling to
the fermions as $\sim g\sigma  \bar\psi^j\psi^j$ and $\sim i g\pi 
 \bar\psi^j\gamma_5 \psi^j$. Gross and Neveu [\GN] computed the effective
potential for these scalar fields. To leading order in ${1\over N}$,
the effective potential for $\sigma$ (at $\pi=0$ e.g.) is given by all
fermion one-loop diagrams with an arbitrary number of 
$\sigma$'s attached. Each individual
diagram is IR divergent, but summing them up  leads, after renormalisation, to
$V_{\rm eff}(\sigma) = {1\over 2}\sigma^2
+{1\over 4\pi} g^2 N \sigma^2\left[\ln\left({\sigma\over \sigma^*}
\right)-3\right]$, where $\sigma^*$ is some fixed subtraction value. This effective
potential has a non-trivial minimum at non-zero $\sigma$. Hence, $\sigma$ develops a
non-vanishing expectation value, $<\sigma>=\sigma^* e^{1-\pi/Ng^2}$ 
which in turn, through
the $\sigma\bar\psi\psi$-interaction, induces a mass $M\sim e^{-\pi/Ng^2}$ for the
fermions. Clearly, this 
dynamical mass generation is non-perturbative.

An alternative way to understand the mass generation is via the 
Bethe ansatz which, of
course, also proves the integrability of the model [\AL]. Let me very briefly
sketch the procedure. One starts with the Hamiltonian that corresponds to the action
\gnaction. This Hamiltonian is then diagonalized in a $n$-particle Fock space. At
this point there is {\it no}  Dirac sea. The diagonalisation of the Hamiltonian
provides the energy levels of the empty Dirac sea (cut off at some large negative
energy since $n$ is finite). The eigenstates are pseudoparticles and correspond to the
fermions of \gnaction\ with zero mass. One has a many-body theory of massless
pseudoparticles. Then one fills this (cut-off) Dirac sea with the pseudoparticles up to
some Fermi level. The physical spectrum  is given by particle-hole type
excitations. Suppose one takes one of the pseudoparticles out of the filled sea. This
creates a hole. But in addition, all other pseudoparticles in the sea are affected by
this hole creation. One can show that all ${n\over 2}$ right-moving pseudoparticles
have their energy and momentum changed by $\Delta E=\Delta k\sim{a\over n}$, while all
left-moving ones suffer a change $\Delta E=-\Delta k\sim{b\over n}$ for some constants
$a, b$. Parametrizing the latter as $a=Me^\theta$ and $b=Me^{-\theta}$
the total change then is
$$\eqalign{
\Delta E_{\rm tot}&={n\over 2}\left( {a\over n}+{b\over n}\right)=M\cosh\theta\cr
\Delta k_{\rm tot}&={n\over 2}\left( {a\over n}-{b\over n}\right)=M\sinh\theta \ .\cr}
\eqn\ii           
$$
This is precisely the dispersion relation for a massive excitation.\foot{
Of course, there is also the hole energy and momentum which still obey a massless
dispersion relation and which decouple [\LOW].} 

Using the Bethe
ansatz, it has been shown[\AL,\LOW] that the $S$-matrix
for pseudoparticle scattering
is factorisable and elastic. Factorisability means that any $m\rightarrow l$ scattering
process can be described by products of $2\rightarrow 2$ $S$-matrix elements.
Elasticity means that in any such scattering process only internal quantum numbers
(e.g. colour) are exchanged between the particles while they keep their individual
momenta and energies. Since physical excitations are made up from pseudoparticle
configurations, it could also be shown [\AL] that the $S$-matrix for the scattering
of the physical excitations, i.e. of the physical (massive) particles, 
also is elastic and factorisable, as a consequence of these same properties for the
pseudoparticle $S$-matrix.

\chapter{Coupling to gravity}

To couple the Gross-Neveu model to two-dimensional gravity, we replace the action
\gnaction\ by its generally covariant form\foot{
Our conventions are: $x^\pm={1\over \sqrt{2}}(x^0\pm x^1),\ \d_\pm=
{1\over \sqrt{2}}(\d_0\pm\d_1)$, $x_ay^a=x_+y^++x_-y^-$. The
Minkowski metric is $\eta_{00}=-1,\, \eta_{11}=1$ $ \Rightarrow\, 
\eta_{+-}=\eta^{+-}=-1$, and $g_{\m\n}=e_{a\m}e^a_{\phantom{a}\n}$,
and $e_a^{\phantom{a}\m}e^a_{\phantom{a}\n}=\delta^\m_\n$ where, as
usual, Lorentz indices ($a,b,\ldots$) are raised and lowered with
$\eta_{ab}$ while $g_{\m\n}$ is used for curved space indices
($\m,\n,\ldots$). One defines $\bar\chi =\chi^+\g^0=\chi\g^0$ and
$\g^0\g^1=\g_5$ so that $\bar\chi\g^0=-\chi$ and
$\bar\chi\g^1=\chi\g_5$. Furthermore $(\g^0)^+=-\g^0, (\g^1)^+=\g^1,
\g_5^+=\g_5$ and $\chi_\pm={1\over 2} (1\pm\g_5)\chi$, so that
$\bar\chi\g^\pm={1\over
\sqrt{2}}\bar\chi(\g^0\pm\g^1)=-\sqrt{2}\chi_\mp$. Finally note that
for a $2\times 2$-matrix $M$ one has $\det
M=M_{00}M_{11}-M_{01}M_{10}=M_{++}M_{--}-M_{+-}M_{-+}$.
}
$$
S_{\rm M}={1\over \sqrt{2}}\int {\rm d}t{\rm d}x (\det e) 
\left[-i\bar\psi^j \gamma^a e_{a}^{\phantom{a}\mu}
\left(\partial_\mu+{1\over 4}\omega_\mu^{\phantom{\mu}cd}\g_{cd}\right)
\psi^j 
+{1\over 2} g^2  (\bar\psi^j\psi^j)^2 - {1\over 2} g^2 (\bar\psi^j\gamma_5\psi^j)^2\right]
\eqn\iii
$$
where $e_{a}^{\phantom{a}\mu}$ is the inverse zweibein, 
$\omega_\mu^{\phantom{\mu}cd}$ the spin-connection 
and $\g_{cd}={1\over 2}[\g_c,\g_d]$. 
It is well-known that for two-dimensional Majorana spinors the spin-connection drops
out. Here however, the $\psi^j$ are Dirac spinors and we need to introduce the 
$\omega_\mu^{\phantom{\mu}cd}$.
We also add the gravity action
$$
S_{\rm G}=-{\gamma\over 16\pi}\int {\rm d}t{\rm d}x \sqrt{-g} R{1\over \nabla^2}R +
\mu \int {\rm d}t{\rm d}x \sqrt{-g} \ .
\eqn\iv
$$
The latter is the well-known induced gravitational action as first derived by Polyakov
[\POLY] (see also section 2 of the second ref. \BK). It can be interpreted as
due to the metric dependence of the ghost and matter measures in the functional
integral. The value of the prefactor $\gamma$ will be given below. 

In conformal gauge the gravitational action \iv\ 
reduces to the Liouville action. Here instead, we will
work in chiral (or light-cone) gauge where the metric and zweibein take the form
$$\eqalign{
{\rm d}s^2&=-2{\rm d}x^+{\rm d}x^--2h_{++}({\rm d}x^+)^2\cr
e_{+-} e_{-+}&=1\ ,\quad e_{--}=0\ ,\quad {e_{++}\over e_{+-}}= h_{++}\ .\cr }
\eqn\v             
$$
We are free to make a convenient choice of the Lorentz phase such that $e_{+-}=-1$.
Then $e_{-+}=-1$, too, and $e_{++}=-h_{++}$. It is then
straightforward to compute the spin-connection from $\omega_{\mu}^{\phantom{\mu}ab}=
-\omega_{\mu}^{\phantom{\mu}ba}$ and
$\d_{[\mu} e_{\nu]}^{\phantom{\mu}a} +\omega_{[\mu}^{\phantom{\mu}ab}
 e_{\nu]}^{\phantom{\mu}c}\eta_{bc}=0$. One readily finds 
$\omega_+^{\phantom{+}+-}=\d_- h_{++}$ and $\omega_-^{\phantom{+}+-}=0$,
so that ${1\over 4}\omega_+^{\phantom{+}cd}\g_{cd}={1\over 2}\d_- h_{++}\g_5$ while 
${1\over 4}\omega_-^{\phantom{+}cd}\g_{cd}=0$.
Finally the matter action takes the form
$$\eqalign{
S_{\rm M}=\int {\rm d}t{\rm d}x 
\Big[&i{\psi^j_-}^*\left( \d_+-h_{++}\d_--{1\over 2} \d_- h_{++} \right) \psi^j_- +
i {\psi^j_+}^* \partial_-\psi^j_+ \cr
&+\sqrt{2} g^2 (\psi_+^i)^* (\psi_-^j)^*\psi_-^i\psi_+^j \Big] \ .\cr}
\eqn\vi
$$
The contribution from the spin connection is of course just what one needs to be able
to rewrite this action in the more symmetric form
$$\eqalign{
S_{\rm M}=\int {\rm d}t{\rm d}x 
\Big[&i{\psi^j_-}^* \d_+  \psi^j_- + i {\psi^j_+}^* \partial_-\psi^j_+
+{i\over 2} h_{++}\left( \d_- {\psi^j_-}^*\psi^j_- -{\psi^j_-}^*\d_- \psi^j_-\right)\cr 
&+\sqrt{2} g^2 (\psi_+^i)^* (\psi_-^j)^*\psi_-^i\psi_+^j \Big] \ .\cr }
\eqn\vii
$$
The gravity action \iv\ in chiral gauge becomes
$$
S_{\rm G}={\gamma\over 8\pi}\int {\rm d}t{\rm d}x
(\partial_-^2 h_{++}){1\over
\partial_-(\partial_+- h_{++}\partial_-)}\partial_-^2 h_{++} 
+\mu \int {\rm d}t{\rm d}x \ .
\eqn\viii
$$
From these expressions one sees that the
matter action splits into a
$h_{++}$-independent part that is exactly the same as in eq.
\gnaction\ plus a term
$\sim   h_{++}\left( \d_- {\psi^j_-}^*  \psi^j_- -{\psi^j_-}^* \d_- \psi^j_-\right)$ 
which
provides an interaction between the right-moving fermions and the 
``graviton" ($h_{++}$).
The left-moving fermions $ \psi^j_+$ do not interact with gravity.
In the gravitational action the second term does not depend on $h_{++}$ and hence does
not  contribute to the dynamics. So we will drop it in the following. The first term,
however, does depend on $h_{++}$ in a non-polynomial way. If we use ${1\over \gamma}$ as
a formal expansion parameter (i.e. $h_{++}\sim {1\over \sqrt{\gamma}}$) one can expand
the $(\partial_+- h_{++}\partial_-)^{-1}$ in a power series leading to
$$
S_{\rm G}={\gamma\over 8\pi}\int {\rm d}t{\rm d}x
 h_{++}\partial_-{1\over\partial_+}
\left[ 1+ \partial_+^{-1} h_{++}\partial_- + (\partial_+^{-1} h_{++}\partial_-)^2 +
\ldots \right]
\partial_-^2 h_{++}  \ .
\eqn\ix
$$
from which one can read of the $h_{++}$ (``graviton")-propagator, as well as the
various vertices involving an arbitrary number of ``gravitons".

The graviton propagator is given by
$$
<h_{++}(-k)h_{++}(k)>=-{4\pi i\over \g}{k_+\over k_-^3}
\eqn\x
$$
with an appropriate $i\epsilon$-prescription to be discussed below, while the vertex
e.g. between three gravitons of momenta $p, q$ and $k=-p-q$ is
$$
 i{\g\over 4\pi} \left[ p^2q^2+p^2k^2+q^2k^2\right] \ .
\eqn\xi
$$
From the matter action \vii\ one reads the fermion propagators ${i\over k_-}$
for the left-moving fermion ($\psi_+$) and ${i\over k_+}$
for the right-moving fermion ($\psi_-$). Most important for us here is the vertex
between an (incoming) right-moving fermion of momentum $p$, an (outgoing) right-moving
fermion of momentum $p'$ and an (incoming) graviton of momentum $k=p'-p$ which is
$$
-{i\over 2}(p_- + p_-') \ .
\eqn\xii
$$
Of course, there is also the four-fermion vertex that was already there before coupling
to gravity. It is $i\sqrt{2}g^2$ between an (incoming) left-moving fermion of colour
$j$,
an (incoming) right-moving fermion of colour $i$ and 
an (outgoing) left-moving fermion of colour $i$ and
an (outgoing) right-moving fermion of colour $j$.

\chapter{The vanishing of the gravitational corrections to the $S$-matrix}

Our goal now is to compute the two pseudoparticle $S$-matrix in the
presence of gravity and show that it is still elastic. As explained above,
the pseudoparticles are the massless fermions described by the action \vii, while
the physical particles are the massive fermions. The pseudoparticles correspond to an
{\it empty} Dirac sea. This means that one cannot create a
pseudoparticle-antipseudoparticle pair. Accordingly, the pseudoparticle propagators
${i\over k_\pm}$ read from \vii\ do {\it not} have the standard Feynman
$i\epsilon$-prescription which would be ${ik_\mp\over k_+k_-+i\epsilon}$. Instead, we
must only use the retarded propagator
$$
<{\psi^l_\mp}^*(-k)\psi^m_\mp(k)>\ =\ \delta_{lm}{i\over k_\pm+i\epsilon}\ =\ 
\delta_{lm}{i k_\mp\over k_+k_- +i\epsilon\, {\rm sgn} k_\mp} \ .
\eqn\xiii
$$
An important consequence of the appearence of the retarded fermion propagators
only, and of the structure of the four-fermion vertex, is that the diagrams of Fig. 1 do
contribute to pseudoparticle scattering amplitudes, while those of Figs. 2 and 3 do not
exist or give vanishing contribution. 

\fig{Diagrams contributing to the left-right scattering}{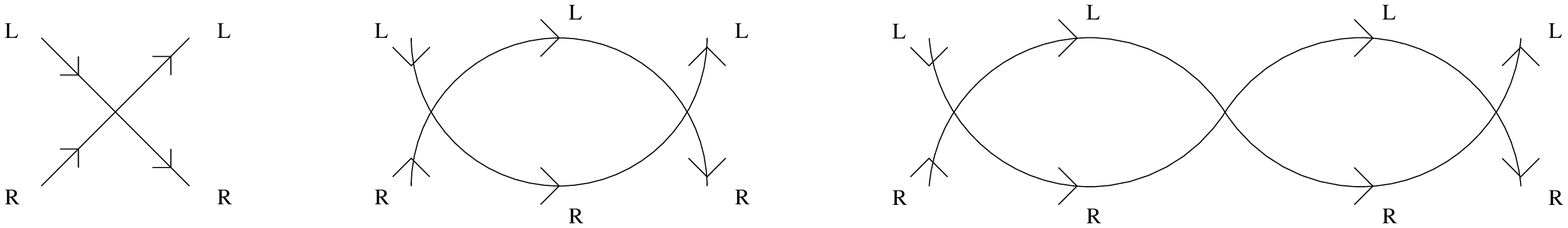}{13cm}
\figlabel\figi
\vskip 2.mm
\fig{Left-left and right-right scattering diagrams giving vanishing contributions}
{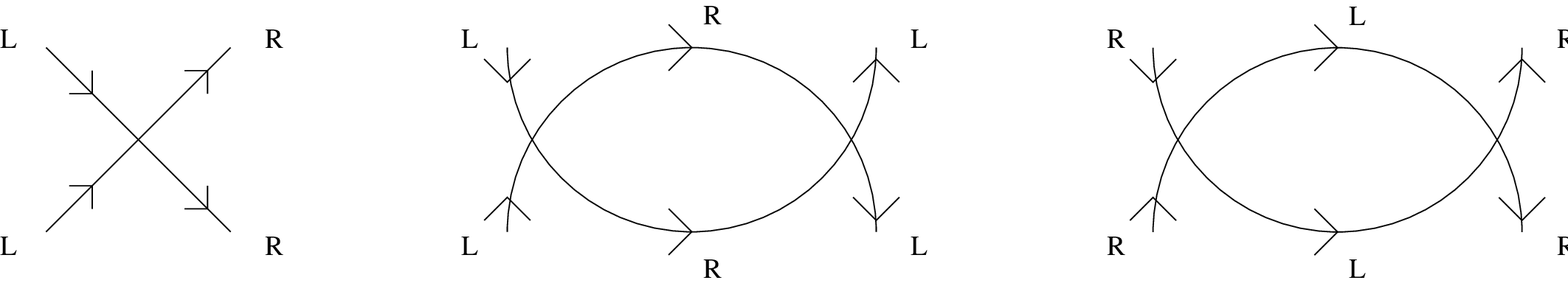}{13cm}
\vskip 2.mm
\fig{Another left-left scattering diagram giving a vanishing contribution}{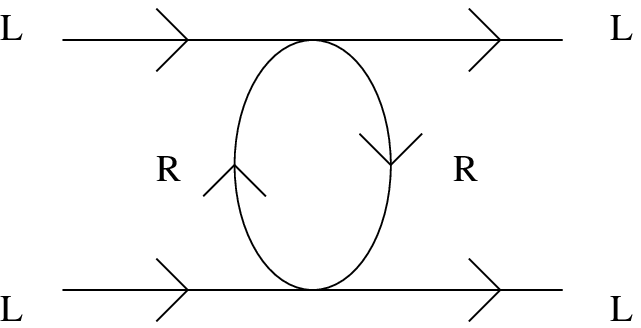}
{5cm}
\vskip 2.mm

Thus, without gravity, there is no left-left
and no right-right scattering. The only non-trivial
scattering is left-right $\rightarrow$ left-right, corresponding to diagrams as in Fig.
1. How does the latter manage to be elastic, i.e. preserve the individual momenta? This
is a simple consequence of energy-momentum conservation and the mass-shell condition.
Indeed, let the initial momenta be $p$ and $q$ and the final momenta $p'$ and $q'$ with
$p,p'$ for the left-moving and $q,q'$ for the right-moving fermions. The conservation
condition then is
$$
p_++q_+=p'_++q'_+\ , \quad p_-+q_-=p'_-+q'_-\ .
\eqn\xiv
$$
The mass-shell conditions are
$$
p_-=p'_-=0\ , \quad q_+=q'_+=0\ .
\eqn\xv
$$
Combining both equations gives
$$
p_+=p'_+\ , \quad q_-=q'_-
\eqn\xvi
$$
which together with \xv\ shows that the individual momenta are unchanged and the
two pseudoparticle $S$-matrix is elastic in the left-right $\rightarrow$ left-right
chanel. Of course, this applies only to the
$S$-matrix; general off-shell amplitudes do not preserve the individual momenta.

\fig{Disconnected diagram contributing only to the gravitational
self-energy of the right fermions}{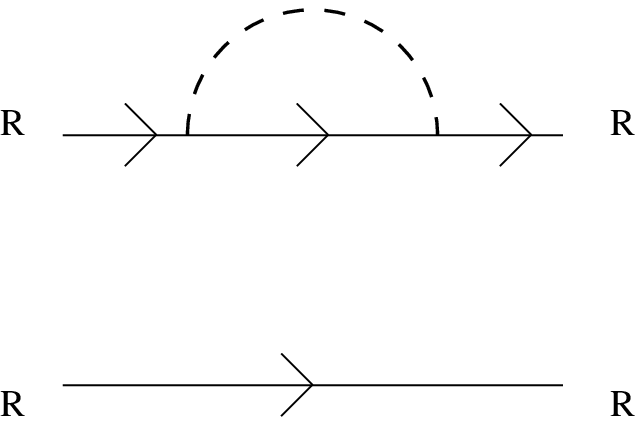}{5cm}
\vskip 2.mm
\fig{Gravitational tree-level scattering of two right fermions}{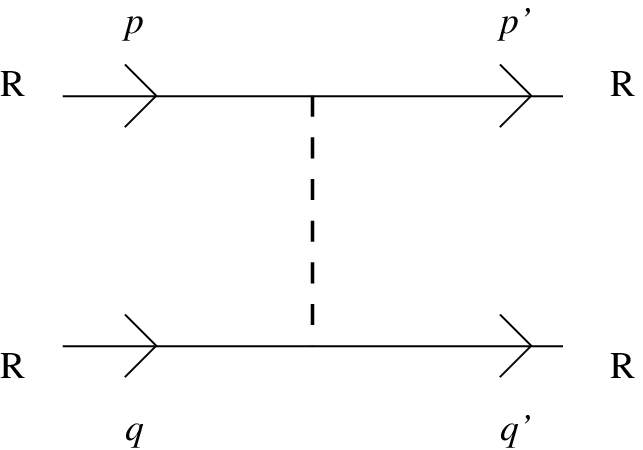}{5cm}
\vskip 2.mm

What changes when we couple this system to gravity? Now the right-moving fermions couple
to the graviton with the vertex \xii. This will affect the
left-right $\rightarrow$ left-right correlation functions, but by the above kinematic
argument the corresponding $S$-matrix element will remain elastic. Also, the
left-left $\rightarrow$ left-left scattering is insensitive to gravity and thus remains
elastic, too.
The only non-trivial case is the right-right $\rightarrow$ right-right 
scattering. There is
no kinematic reason for this to remain elastic, and the fermions do couple to gravity
via diagrams as shown in Figs. 4 - 7. Of course, the diagram of Fig. 4 cannot lead to
any non-trivial scattering. The first diagram that could do so is the tree-diagram of
Fig. 5 involving a one-graviton exchange between two right-moving fermions. 
Recall from eq. \x\ that the graviton propagator is $-{4\pi i\over
\gamma} {k_+\over k_-^3}$. Contrary to the fermions, it should be the Feynman
propagator and have the standard $i\epsilon$-prescription, i.e.
$$
-{4\pi i\over\gamma} {k_+\over k_-^3} \ \rightarrow \ 
-{4\pi i\over\gamma} {k_+^4\over (k_+ k_-+i\epsilon)^3}
= -{4\pi i\over\gamma} {k_+\over (k_-+i\epsilon \, {\rm sgn}k_+)^3} \ .
\eqn\xvii
$$

Let's first compute the diagram of Fig. 5. Actually, we will compute the amputated
diagram on shell as relevant for the $S$-matrix. It is given by
$$
{i\pi\over \g} (p_-+p_-')(q_-+q_-'){(p_+-p_+')\over (p_--p_-')^3} \ .
\eqn\xviii
$$
Putting the external momenta on shell, $p_+=p_+'=q_+=q_+'=0$ for right-moving fermions,
the amplitude \xviii\ vanishes. It is clear from the structure of the graviton
propagator that one always gets zero as long as it is coupled to two on-shell
right-moving fermions. It then immediately follows that both diagrams of Fig. 6 also
vanish on shell.
\fig{Vertex corrections to the diagram of Fig. 5}{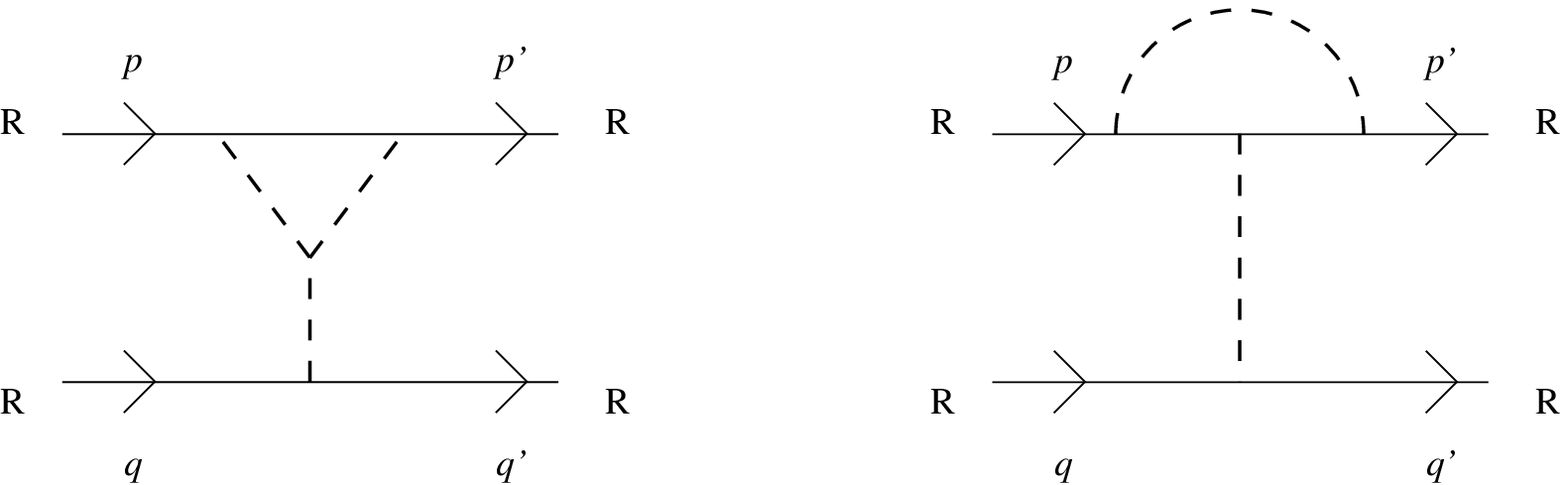}{12cm}
\vskip 2.mm
\fig{The box and crossed box diagrams contributing to the one-loop gravitational
scattering of two right fermions}{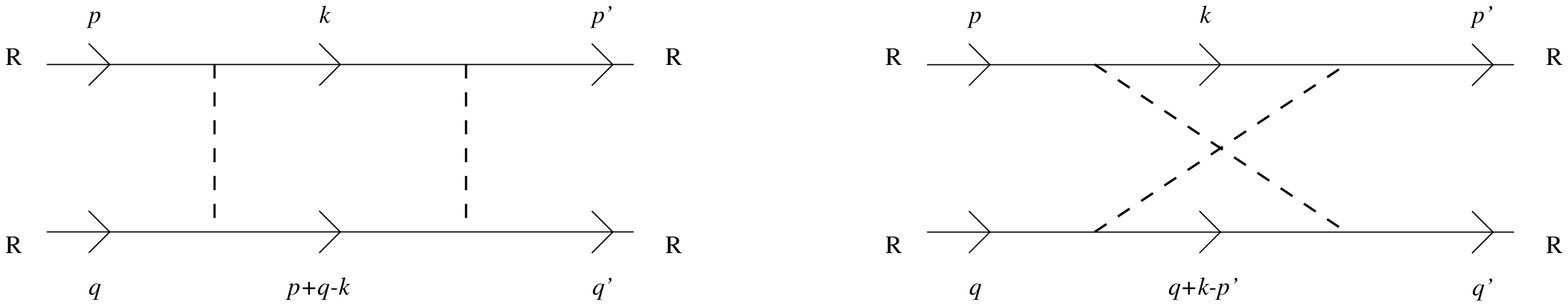}{14cm}
\vskip 2.mm
So, up to order ${1\over \g^2}$, we are left with the two diagrams of
Fig. 7 only. They yield
$$\eqalign{
& {1\over 4\g^2}\int {\rm d}^2k \ 
{(p_-+k_-)(p_-'+k_-)(p_+-k_+)(k_+-p_+')\over k_+(p_--k_-)^3(k_--p_-')^3 } \times \cr
& \times \left[ {(2q_-+p_--k_-)(2q_-'+p_-'-k_-)\over q_++p_+-k_+}
+{(2q_--p_-'+k_-)(2q_-'-p_-+k_-)\over q_+-p_+'+k_+}\right] .\cr}
\eqn\xix
$$
Again, we only want to evaluate it on-shell where it reduces to
$$
-{(q_-+q_-')\over 2\g^2}\int {\rm d}^2k \ 
{(k_-+p_-)(k_-+p_-')(2k_--p_--p_-')\over (k_--p_-)^3(k_--p_-')^3 }  \ .
\eqn\xx
$$
Although it looks as if the integrand does not depend on $k_+$, this is not true.
The $k_+$-dependence comes in through the $i\epsilon$-term in the graviton propagators.
Let's integrate over $k_-$ first. The integrand falls off fast enough at infinity so
that one can close the contour by a semicircle in the upper or lower half-plane. We
have two third-order poles (assuming $p_-\ne p_-'$ i.e. $p\ne p'$; if $p=p'$ we have
a pole of order six and the integral obviously vanishes). 
According to the $i\epsilon$-prescription \xvii\ they
are at $k_-=p_- -i\epsilon\, {\rm sgn} (k_+-p_+)$ and
$k_-=p_-' -i\epsilon\, {\rm sgn} (k_+-p_+')$. If we choose the semicircle in the lower
half-plane, we will pick up residues from  the first pole only if $k_+>p_+$
and from the second pole only if
$k_+>p_+'$. One gets
$$\eqalign{
& - {2\pi i\over \g^2} {(q_-+q_-')(p_-+p_-')\over (p_--p_-')^3} 
\int {\rm d}^2k \left[ \theta(k_+-p_+)-\theta(k_+-p_+')\right] \cr
& =  {2\pi i\over \g^2} {(q_-+q_-')(p_-+p_-')(p_+-p_+')\over (p_--p_-')^3} \ .\cr}
\eqn\xxin
$$
Had we chosen the semi-circle in the upper half-plane, the result would have been the
same, of course.
Thus the one-loop result (on shell) is, up to the factor ${2\over \g}$, identical to
the tree-level result, and hence vanishes for the same reason ($p_+=p_+'=0$ on shell).

Let me stress that the vanishing of the scattering amplitude at one loop was in no way
obvious a priori. One has to add both diagrams of Fig. 7 and use the on shell condition.
Then, it is only through the subtlety of the  $i\epsilon$-prescription  that one gets
the factor $(p_+-p_+')$ which makes the amplitude vanish on shell. It is very tempting
to speculate that this sort of mechanism will persist at all orders in ${1\over \g}$.

\chapter{Conclusions}

We have seen that, in general, 
the coupling to gravity does modify the right-right $\rightarrow$ right-right 
correlation functions. However, the gravitational corrections to the {\it on shell}
right-right $\rightarrow$ right-right
scattering amplitude vanish at tree and one-loop level, i.e. up to and including order
${1\over \g^2}$. We thus conjectured [\BK] that this might be true to all orders in
${1\over \g}$. Let's assume here that this is indeed the case, and see what it implies.

First of all, one might object the use of perturbation theory in ${1\over \g}$. The
constant $\g$ is given by [\POLY]
$$
\gamma={1\over 12}\left( c-13-\sqrt{(c-1)(c-25)}\right)
$$
where $c$ is the total matter central charge. Of course, gravity is well understood for
$c\le 1$ where $\gamma$ is real. This is the weak coupling regime. Indeed, as
$c\rightarrow -\infty$ one has $\gamma\sim{c\over 6}$. Since ${1\over \gamma}$ is the
gravitational coupling constant, $c\rightarrow -\infty$ is the gravitational
weak-coupling limit. A perturbation
expansion in ${1\over \gamma}$ could be expected to be
reasonable as long as $c$ is large and
negative. In the Gross-Neveu model this is certainly not the case. Moreover, in ref.
\BK\ we have shown that, in the presence of gravity,
certain fermion correlation functions  have a diverging expansion in ${1\over \g}$ for
all $\g$, and that the Borel-resummed perturbation series exhibits a typically
non-perturbative behaviour. Here, however, the situation is different: the perturbative
expansion of the right-right $\rightarrow$ right-right on shell scattering amplitude, 
having only zero coefficients, converges everywhere in the complex
${1\over\g}$-plane, yielding zero for all $\g$. 
Although one cannot exclude a non-perturbative
contribution $\sim e^{-a\g}$ to the $S$-matrix, this does not seem to be very likely.

So if the gravitational corrections to the
right-right $\rightarrow$ right-right (pseudoparticle) $S$-matrix
elements do indeed vanish, as they do for the left-left $\rightarrow$ left-left
elements, the only corrections are to the left-right $\rightarrow$ left-right
(pseudoparticle) $S$-matrix
elements. But as already noted earlier, the latter nevertheless remain elastic, and
one would conclude that the two pseudoparticle scattering $S$-matrix remains
elastic in all chanels when the coupling to gravity is included. 
One could then go on and speculate
that all $S$-matrix elements for multi-pseudoparticle scattering factorize, and hence
reduce to products of two-pseudoparticle scattering $S$-matrices, which are all
elastic. Thus the complete $S$-matrix for pseudoparticle scattering would be
factorisable and elastic. Since the physical $S$-matrix for the scattering of the
physical (massive) fermions is obtained from the pseudoparticle $S$-matrix, it
probably would turn out to be elastic and factorisable, too. If this chain of
hypothesis goes through, the chiral Gross-Neveu model would indeed remain
integrable when coupled to gravity. However, there is still a long way to go.

\ack
It is a great pleasure to thank Ian Kogan for the enjoyable collaboration that led to
the results presented here. We also had profited from discussions with D. Gross and A.
Polyakov.

\vskip 5.mm
\noindent
$\underline{\rm Note\ Added:}$ Let me note that after our original papers [\BK] and 
at about 
the same time as my lecture at the Cargese summer school, 
there appeared a paper by Abdalla and
Abdalla [\AA] where the gravitational dressing of integrable models was studied in
conformal gauge. In particular, for the Gross-Neveu model they claimed that coupling to
gravity does not invalidate the existence of higher conservation laws, and they
concluded that the model remained integrable, thus confirming our conjecture.

\refout

\end